\documentclass[letterpaper, 10 pt, conference]{ieeeconf}  % Comment this line out if you need a4paper

\IEEEoverridecommandlockouts                              % This command is only needed if 
\usepackage{graphicx}
\usepackage{xcolor}
\usepackage{svg}
\title{\LARGE \bf
Edge Dynamic Map architecture for C-ITS applications
}
\author{Mikel García$^{1,2}$, Gorka Vélez$^{1}$, Josu Pérez$^{1}$, Ángel Martín$^{1}$, Zaloa Fernández$^{1}$ and Naiara Aginako$^{2}$% <-this % stops a space
\thanks{Funded by the European Union, under Horizon 2020 research and innovation programme (project 5GMETA, grant agreement 957360).}
\thanks{$^{1}$ Vicomtech Foundation, Basque Research and Technology Alliance (BRTA), Mikeletegi 57, 20009, Donostia-San Sebasti\'{a}n (Spain).} 
\thanks{$^{2}$ University of the Basque Country (UPV/EHU), Donostia-San Sebasti\'{a}n (Spain).}
}

\begin{document}

\maketitle
\thispagestyle{empty}
\pagestyle{empty}

%%%%%%%%%%%%%%%%%%%%%%%%%%%%%%%%%%%%%%%%%%%%%%%%%%%%%%%%%%%%%%%%%%%%%%%%%%%%%%%%
\begin{abstract}
Cooperative Intelligent Transport Systems (C-ITS) create, share and process massive amounts of data which needs to be real-time managed to enable new cooperative and autonomous driving applications. Vehicle-to-Everything (V2X) communications facilitate information exchange among vehicles and infrastructures using various protocols. By providing computer power, data storage, and low latency capabilities, Multi-access Edge Computing (MEC) has become a key enabling technology in the transport industry. The Local Dynamic Map (LDM) concept has consequently been extended to its utilisation in MECs, into an efficient, collaborative, and centralised Edge Dynamic Map (EDM) for C-ITS applications. This research presents an EDM architecture for V2X communications and implements a real-time proof-of-concept using a Time-Series Database (TSDB) engine to store vehicular message information. The performance evaluation includes data insertion and querying, assessing the system's capacity and scale for low-latency Cooperative Awareness Message (CAM) applications. Traffic simulations using SUMO have been employed to generate virtual routes for thousands of vehicles, demonstrating the transmission of virtual CAM messages to the EDM.
\end{abstract}

\section{Introduction}
\label{sec:introduction}

Recent developments in the automotive and telecommunications industries have garnered significant interest in implementing Vehicle-to-Everything (V2X) standards and Cooperative Intelligent Transport Systems (C-ITS). These technologies facilitate communication between vehicles, surrounding infrastructure, and among vehicles themselves. This enables the development of new cooperative driving and information services that would not be possible without V2X and C-ITS, thereby enhancing the efficiency, safety, and sustainability of transportation systems. Thanks to V2X technologies, road users can share messages using standards such as ETSI Cooperative Awareness Message \cite{ETSI2019} (CAM) to send information about each other's dynamics, position and attributes or ETSI Collective Perception Messages \cite{ETSI2019_2} (CPMs) adding information about road users or obstacles detected by an on-board perception system. The standards establish the format and set of rules for the generation frequency of messages, which vary according to factors such as road user dynamics, manoeuvres, or other considerations.

On Board Units (OBU) can generate, send and receive these standardised messages using 5G networks or IEEE 802.11p, the basis of the DSRC (Dedicated Short Range Communications) and ITS-G5 \cite{9133075}. However, there are no standardised ways to manage (fuse, align, and integrate) vehicular messages between different traffic participants. A Local Dynamic Map (LDM) serves as a local and centralised data structure generated from heterogeneous data sources, including onboard sensors and V2X messages. The LDM concept was originally introduced in the SAFESPOT project, proposing a four-layer structure based on the level of dynamicity of the stored objects \cite{Andreone2010}. Several models extending base static maps have been proposed for the categorisation of the environment, including a five-layer structure used by Lyft \cite{Muñoz2022}, and a six-layer structure proposed in \cite{Scholtes2021}. In any case, there is no established method for implementing an LDM data structure. Some authors, including the seminal SAFESPOT project, proposed the use of a SQL database \cite{Shimada2015, Vilalta2020}, for instance, using PostgreSQL and PostGIS as a library for spatial operation extending PostgreSQL \cite{Shimada2015}. In the LDM approach proposed in \cite{Eiter2019}, the data received from V2X messages is added either to the static database or to the stream database implemented by PipelineDB. In \cite{Puphal2022}, a relational Local Dynamic Map (R-LDM) embodied as an interconnected graph of nodes is used for risk estimation. A similar approach is used in \cite{Garcia2022}, implementing a graph database with a focus on interoperability with OpenLABEL as a common data format. Inspired by the concept of LDMs, a vehicular video streaming solution that keeps a history log of all data collected by the cameras is proposed in \cite{Maiouak2019,Maiouak20192}.

In this work, we propose to extend the concept of LDM to Edge Dynamic Map (EDM), to fuse V2X communications, exploiting the ultra-low latency of 5G networks, and the capabilities of Multi-access Edge Computing (MEC) computing. MEC brings data storage and computing power capabilities closer to the edge network, allowing lower latency and faster processing in applications that require real-time communication and decision-making. These features make MEC technology suitable for several vehicular applications \cite{Ling2022}. The traditional V2X approach on ad-hoc networks is based on broadcast messages: vehicles send the same message to the rest of the vehicles, which are all listening. So each vehicle needs to construct its own LDM based on its own perception and the messages that it receives from surrounding vehicles. In this paper, we propose  a centralised server deployed in a MEC, that collects C-ITS messages from connected vehicles and builds up an EDM. This EDM can be accessed by any connected vehicle or by an application or service deployed in the MEC. 

The proposed EDM follows the structure of the original four-layer implementation; however, this work will focus on the static and dynamic layer using a Time-Series Database (TSDB). The TSDB contains the dynamic information captured by the connected vehicles and received through C-ITS messages. We propose using a TSDB as it is more suited to the time-dependent nature of the dynamic objects. In a TSDB, old data can efficiently be removed from the database, and data can be queried for a specific time span. This work builds upon the work presented in \cite{velez2022}, extending and improving several aspects: 1) adding the static layer and creating the EDM concept; 2) making the proposed method suitable for a variety of ITS use cases instead of being dedicated only to a vehicle discovery service; 3) adapting the architecture to address the challenge of node mobility \cite{Fernandez2023}; 4), adding geolocation-based vehicle filtering using a geospatial indexing library; 5) improving the insertion of data to the database using data buffering and batch insertion; and 6) studying the performance and scalability of the new solution. The rest of the paper is organised as follows: Section \ref{sec:arch} describes the architecture of the proposed EDM, Section \ref{sec:implementation} describes the implementation of the architecture, Section \ref{sec:res} shows the results obtained in the experimentation, and Section \ref{sec:conclusions} concludes the paper.

%--------------------------------------------------------------------------------------------------------
% -------------------------------------------ARCHITECTURE------------------------------------------------
%--------------------------------------------------------------------------------------------------------

\section{Architecture}\label{sec:arch}

\begin{figure}[!ht]
    \centering
    \includegraphics[width=0.48\textwidth]{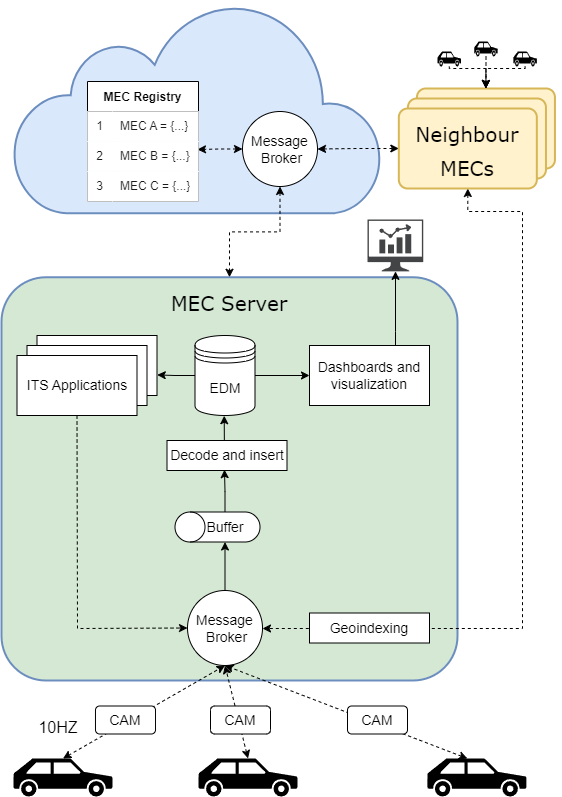}
    \caption{High-level representation of the proposed architecture.}
    \label{fig:Architecture}
\end{figure}

The proposed EDM architecture is presented in Fig. \ref{fig:Architecture}. The system uses message brokers to facilitate communication by implementing a publish-subscribe protocol. This protocol uses messaging topics, which are hierarchical in nature, and allow the dissemination of messages from a publisher to multiple subscribers. Our architecture is defined by three main components: the MEC registry, the MEC server and the client side. In addition, the messaging broker and the MEC interconnection strategy are also detailed. 

\subsection{MEC Server}
The MEC server comprises three main components: a message broker, an EDM, and ITS applications. All the messages sent by the vehicles are received in the message broker. The EDM, composed of a database and a static map of its operational area, consumes the information received in the message broker and inserts it into its local database. A temporal threshold can be applied to delete or store old data from the database, depending on the ITS application's temporal window needs. ITS applications can be built to interact with the information stored in the EDM directly. Each MEC server is defined by attributes that provide relevant information to clients and other MEC servers. These attributes include the MEC server's position in WGS-84 coordinates, an MEC ID, its operational and optimal coverage areas, and messaging clients connected to the MEC registry and neighbouring MEC servers. In our architecture, the MEC servers’ coverage areas can overlap with respect to other MEC servers. MEC servers need to be aware of their neighbour MEC servers to obtain relevant vehicle information in border areas and provide handover information to clients if they need to switch connection to a neighbour MEC.

\subsection{MEC Registry}
The MEC registry is responsible for maintaining an updated copy of the state of all the MEC servers in a given area. A reference MEC registry address must be provided every time a new MEC server is deployed. When connecting for the first time, the MEC registry will store the information of the MEC server and forward the information of neighbour MEC servers based on its geolocation while updating the new neighbour to the other existing MEC servers. If any of the MEC server attributes changes (position, disconnection, coverage areas), the MEC registry is notified of the change and will update the values of that MEC server and forward the information to its neighbour MEC servers. The MEC registry is also the reference point for a new client connection; every time a new client is connected, it will send information about its position to the MEC registry. The MEC registry will assign the client a unique ID and calculate the most suitable MEC server in its area based on its geoposition. The MEC registry is, therefore, a key element of MEC discovery.

\subsection{Clients}
Clients will use the standard ETSI CAM to communicate with the message broker. CAMs provide relevant information about a vehicle status including position, heading, speed, acceleration, etc., which allows the system to insert relevant vehicle information into the EDM.
The first time a vehicle client is connected, it must connect to the MEC registry and send a CAM message through a login topic of the message broker. Once this is done, the MEC registry will send the client the address of the most suitable MEC server in its area. As defined in \cite{ETSI2019}, CAM messages shall be sent at a minimum rate of 1 Hz and maximum of 10 Hz. In this work, we consider that client vehicles send CAMs at their maximum send rate of 10 Hz. Clients can directly query the MEC server ITS applications to get relevant information about vehicles in their surroundings. If clients are located in a border area of the coverage of their MEC server, they may receive a handover message from the server with the address information of a neighbouring MEC server that has better network coverage in the area, allowing the client to connect to the new server before a service interruption occurs.
\begin{table*}[]
\centering
\caption{Message Broker Topic architecture proposal. M = MEC, V = Vehicles, N = Neighbour MEC, R = Registry.}
\label{table:mqtt}
\begin{tabular}{cccccl}
\hline
\textbf{Topic name} & \textbf{Broker} & \textbf{Publishers} & \textbf{Subscribers} & \textbf{Triggers} & \multicolumn{1}{c}{\textbf{Topic name}} \\ \hline
\textbf{CAM Feed} & M & V & M, N & No & \textbf{mec\_id/edm\_feed/geo\_index} \\ \hline
\textbf{ITS Query} & M & V & M & Topic Query Response & \textbf{mec\_id/its\_app\_id/query/vehicle\_id} \\ \hline
\textbf{ITS Query Response} & M & M & V & No & \textbf{mec\_id/its\_app\_id/response/vehicle\_id} \\ \hline
\textbf{Vehicle Login} & R & V & R & Topic Login Response & \textbf{mec\_registry\_id/vehicle/login} \\ \hline
\textbf{Vehicle Login response} & R & R & V & No & \textbf{mec\_registry\_id/login\_response/vehicle\_id} \\ \hline
\textbf{Vehicle Handover} & M & M & V & Vehicle changes MEC & \textbf{mec\_id/handover/vehicle\_id} \\ \hline
\textbf{MEC Login} & R & M & R & Topic Neighbour Update & \textbf{mec\_registry\_id/mec/login} \\ \hline
\textbf{MEC Update} & R & M, R & M, R & Topic Neighbour Update & \textbf{mec\_registry/update/mec\_id} \\ \hline
\textbf{Neighbour Update} & R & R & M & No & \textbf{mec\_registry/neighbours/mec\_id} \\ \hline
\end{tabular}
\end{table*}

\subsection{Messaging Broker Topic Architecture}
A proposal for the necessary messaging topics to allow communication between each element in our architecture is summarised in Table I. Some topics will be unique, while others will have subtopics based on geolocation indexes or client IDs to share the information only with
the necessary participants in that communication process. Each of the proposed messaging topics serves a specific objective. The CAM feed will include subtopics composed of geo-position indexes from the received vehicle messages. The EDM will decode and insert CAM messages received in this topic. In the ITS query topic, a given ITS application can be queried, and it will publish its results to its reference ITS query response topic. On startup, vehicles must send a CAM message to the vehicle login topic of the MEC registry, which will forward the most suitable MEC server address using the vehicle login response topic.

Based on the optimal and operating ranges defined for each MEC server and its neighbours, some areas will be identified as handover areas. If a vehicle approaches one of these areas, the MEC server will forward a handover message with the address of the most suitable neighbour MEC server. This handover is triggered following a hysteresis loop to avoid a ping-pong effect. The MEC login and update topics will be the main communication channel between MEC servers and the MEC registry to deploy new MEC servers or update existing ones. When a new MEC server is deployed or updated, the MEC registry will forward its information only to the involved neighbour MEC servers through the neighbour update topic. 
These are the main necessary topics for our architecture; new topics to allow new features can be easily added, allowing an easy implementation for other use cases.

\subsection{MEC Interconnection}
Sharing relevant information between MEC servers is essential to maintain an updated EDM in border regions of their coverage areas. Given the optimal and operating range of each MEC server, we could constantly calculate if each CAM message is inside a neighbour MEC coverage area and forward this information, but this quickly becomes inefficient as the number of vehicles connected to the MEC server grows. To avoid this scalability problem, we propose to organise the broker feed topics into geospatial areas with unique indexes. As the MEC registry shares information about neighbour MEC servers, each MEC server can connect to its neighbour MQTT broker and subscribe only to the relevant CAM feed topic names in their border regions.

%--------------------------------------------------------------------------------------------------------
% -------------------------------------------IMPLEMENTATION----------------------------------------------
%--------------------------------------------------------------------------------------------------------

\section{Implementation}\label{sec:implementation}
Communication across different levels of the proposed architecture is handled using the MQTT \cite{Light2017} messaging protocol designed for the Internet of Things (IoT). MQTT is an extremely lightweight publish-subscribe messaging system also used in automotive applications. The MEC server and the MEC Registry were implemented using Docker and Python3. This section is organised as follows. First, we present our geoindexing approach used to efficiently index and retrieve data from the EDM. Second, we describe the EDM implementation details, including the selected database, message buffering approach and message processing time considerations. Finally, we discuss the visualisation and data monitoring capabilities of our solution.

\subsection{Geoindexing}
\begin{figure}[!ht]
    \centering
    \includegraphics[width=0.48\textwidth]{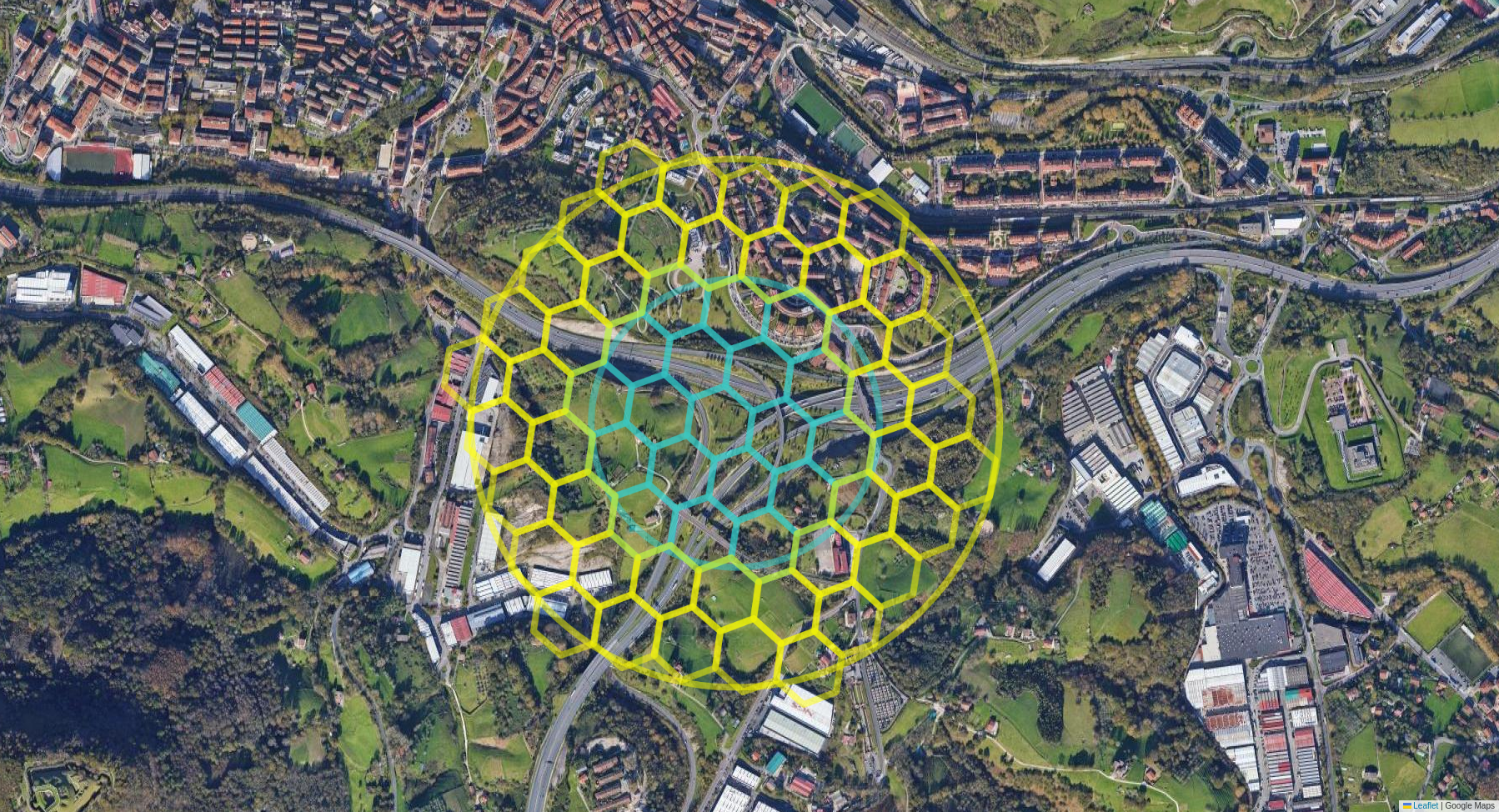}
    \caption{H3 hexagon size for the selected resolution. Yellow hexagons represent indexes within the operating range, while blue hexagons represent those within the optimal range.}
    \label{fig:h3}
\end{figure}

In our MQTT topic architecture, each vehicle needs to publish into the CAM feed using a unique geo-position index based on its location. For this purpose, we are using Uber's hexagonal hierarchical geospatial indexing library H3. H3 facilitates the indexing of geospatial coordinates within hexagons of diverse resolutions. Resolutions greatly modify hexagon sizes, with the biggest hexagon area resolution being $4,250,546$km$^2$ and $0.9$m$^2$ being the smallest. In this study, we selected a resolution with an average hexagon area of $\sim 15,000$m$^2$ (see Fig. \ref{fig:h3}), corresponding to a maximum distance of $\sim130$m between edges of the hexagon, this resolution can be easily modified to match other use case requirements. 
%This resolution lets us experiment with a high number of simulated vehicles to test the scalability of the proposed solution. In the end, the variable that most stresses the EDM is the number of connected vehicles rather than the area that the MEC covers. In any case, the geographical distribution of MECs in a real-world deployment is still a matter of discussion, which depends on a techno-economic analysis \cite{Chiha2023} that is out of the scope of the present work.

\subsection{EDM}
The EDM of the MEC Server has been implemented with InfluxDB, a TSDB, as our database engine to enable CAM data storage and ITS applications querying functions in the EDM. A TSDB efficiently filters data retrieved from queries by selecting a time window threshold. This is essential for real-time applications that only need to access the most recent stored information. The EDM is enhanced with road information of their coverage areas using public and open-source information obtained from Open Street Map files.

For each vehicle publishing CAM messages into the MQTT Broker, the following fields are being stored into the TSDB: \textit{id, timestamp, latitude, longitude, type, heading, speed, acceleration, h3index}. InfluxDB allows to insert data into the database individually or in batches. Performing batch insertions provides better performance than writing data into the TSDB individually. To take advantage of this, we implemented a message buffering system to perform batch write operations in the TSDB and performed an analysis of data insertion times in Section \ref{sec:res}. The purpose of this analysis is to check the number of vehicles that can be handled depending on the buffering time window size set in our solution.

\subsubsection{EDM message buffering}

Messages received in the MQTT Broker are constantly stored into a buffer. As the maximum frequency that a vehicle should be sending V2X messaging is 10 Hz, messages should be available in the database in less than 100ms from the message generation time. We have taken into account the following processing or latency times when measuring an end-to-end latency from a client to the database. 
\begin{equation}
    t_{msg} = t_{send} + t_{buffer} + t_{decode} + t_{insertion}
\end{equation}

Where $t_{send}$ is the sending latency from the client to the MEC server, $t_{buffer}$ is the buffering time window set for the MEC, $t_{decode}$ the time it takes to decode the CAM message and $t_{insertion}$ is the insertion time of the messages into the TSDB. Considering a real-time use case, the following condition should be fulfilled $t_{msg} < 100ms$. In our approach, when the buffering time is completed, a thread is launched to decode and insert the messages into the database. So the following condition should be fulfilled as well $t_{buffer} > t_{decode} + t_{insertion}$, otherwise, the processing thread would not be able to handle the number of messages received. 

\subsection{Data monitoring and visualisation}
\begin{figure}[h]
    \centering
    \includegraphics[width=0.48\textwidth]{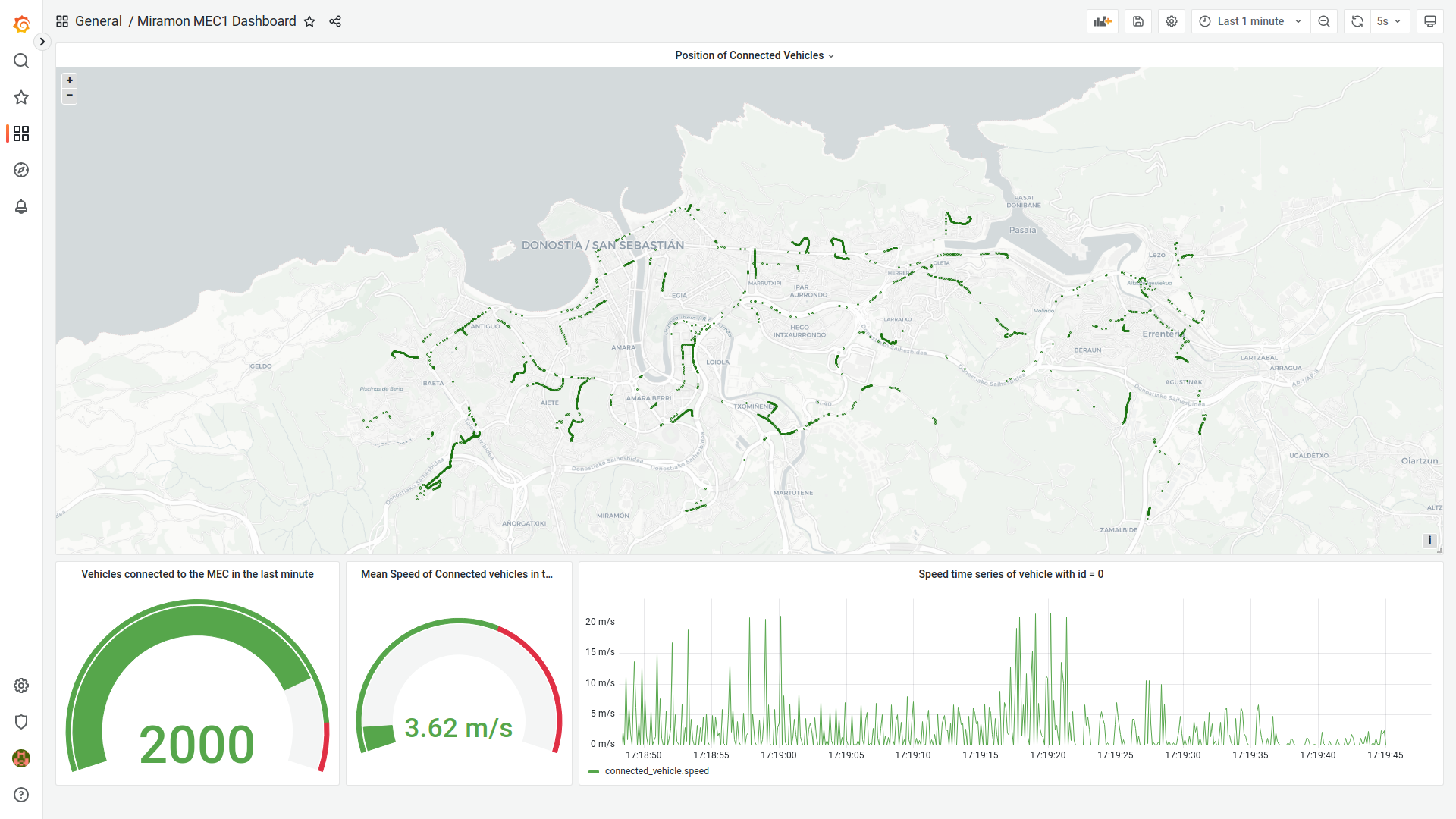}
    \caption{Grafana dashboard visualisation of the data stored in the EDM.}
    \label{fig:grafana}
\end{figure}

Thanks to the usage of InfluxDB, Grafana can be easily connected to the EDM in order to monitor the real-time and historical status of the vehicles connected to the MEC. Grafana allows the creation of dashboards for metrics visualisation and resource management. By applying periodical queries directly from Grafana to the TSDB, data from the EDM can be easily visualised and monitored, as we can see in Fig. \ref{fig:grafana}.

By applying a set of rules and querying the EDM, alerts can be configured using Grafana to notify users or traffic agents of traffic jams, accidents, road works, etc.

%--------------------------------------------------------------------------------------------------------
% -------------------------------------------EXPERIMENTS-------------------------------------------------
%--------------------------------------------------------------------------------------------------------

\section{Results}\label{sec:res}
To evaluate the performance of the proposed architecture, we tested the system's insertion capabilities using a set of virtual CAM messages obtained from traffic route simulations conducted with the SUMO \cite{SUMO2018} simulator. We tested the database performance for insertion of batches of different sizes and querying performance time. The tests have been performed in a server with an Intel Core I5-9400F @2.9GHz processor and 8GB DDR4 RAM in Ubuntu 20.04. This can be seen as a modest computing resource, but it is important to note that the EDM is designed to be deployed in a MEC. In the Edge/Cloud paradigm, it is assumed that the latency-critical computing applications are deployed in the decentralised MEC, while compute-intensive and delay-tolerant applications are deployed in the conventional remote cloud \cite{Lina2022}. So computing resources in the MEC are much more limited than in the Cloud. All the tests have been performed selecting a cache size of 1MB for the TSDB.

\subsection{Data insertion}
\begin{figure}[h]
    \centering
    \includegraphics[width=0.48\textwidth]{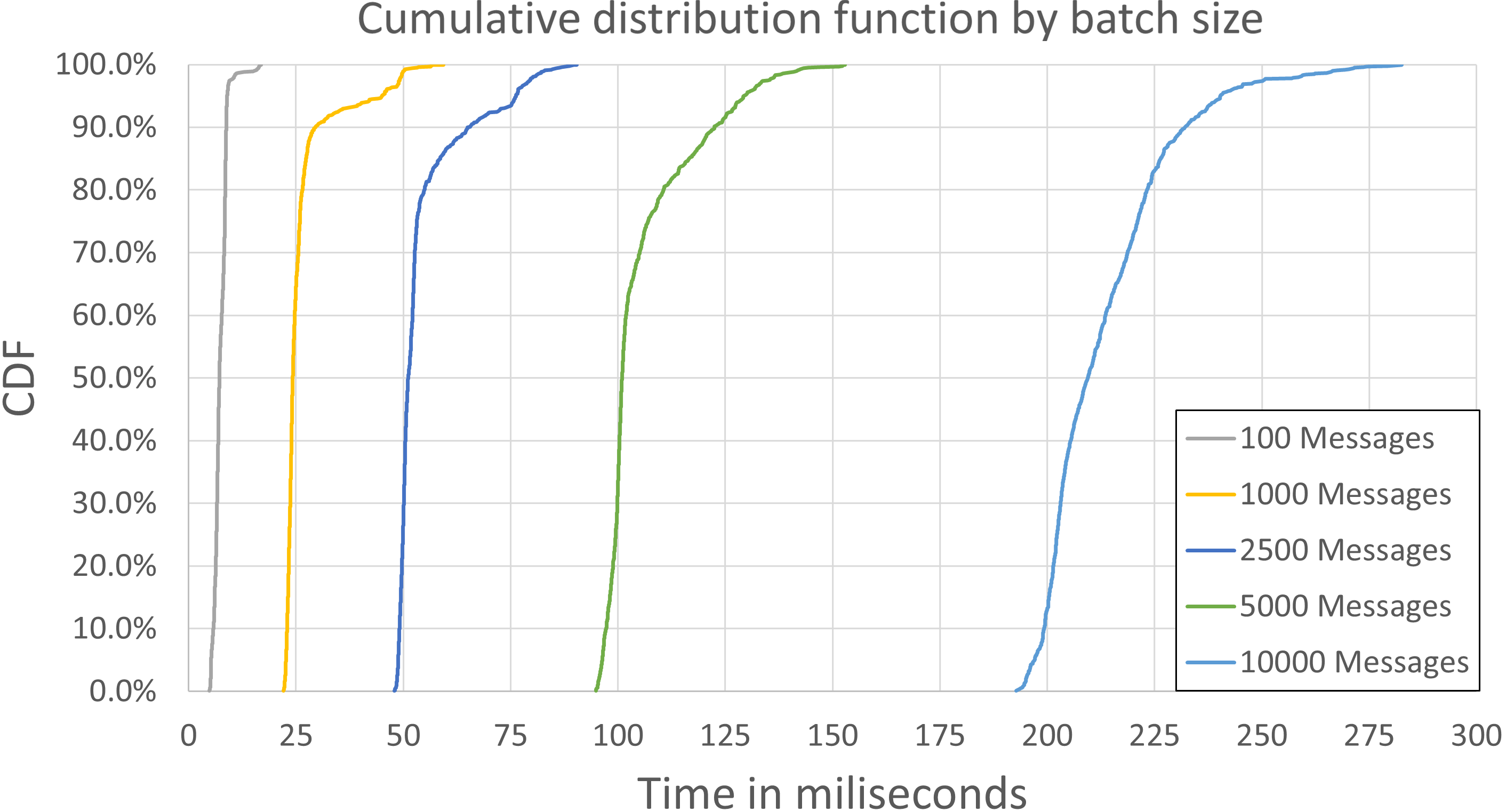}
    %\caption{Cumulative distribution function of the total time for decoding and inserting the messages of each batch size.}
    \caption{Total decoding and insertion time CDF plot per batch size.}
    \label{fig:insertion}
\end{figure}

Data insertion into the database has been analysed for CAM batches of 100, 1000, 2500, 5000 and 10000 messages generated from the SUMO simulations. When measuring this, we have measured both CAM message decoding time (including string formatting to Influx line protocol) and database insertion times. For each batch size, we performed 1000 insertions into the database. Results can be seen in Fig. \ref{fig:insertion} and Table \ref{table:insert_edm}.

\begin{table}[h]
\centering
\caption{Mean (standard deviation) decoding and insertion time in milliseconds for the selected batch sizes.}
\label{table:insert_edm}
\begin{tabular}{l|c|c|}
\cline{2-3}
\multicolumn{1}{c|}{} & \textbf{Decode} & \textbf{Insertion} \\ \hline
\multicolumn{1}{|l|}{\textbf{Batch 100}} & 1.98 (0.61) & 5.44 (1.06) \\ \hline
\multicolumn{1}{|l|}{\textbf{Batch 1000}} & 12.46 (5.02) & 13.92 (3.29) \\ \hline
\multicolumn{1}{|l|}{\textbf{Batch 2500}} & 28.25 (7.00) & 26.02 (4.25) \\ \hline
\multicolumn{1}{|l|}{\textbf{Batch 5000}} & 54.67 (7.31) & 50.94 (7.75) \\ \hline
\multicolumn{1}{|l|}{\textbf{Batch 10000}} & 105.71 (6.61) & 107.74 (14.10) \\ \hline
\end{tabular}
\end{table}

For real-time ITS applications, the requirement is to store sent messages in the databases in less than 100 ms. Taking into account that our message buffering approach needs that $t_{buffer} > t_{decode} + t_{insertion}$, then $t_{decode} + t_{insertion} < 50 ms$ in order to be able to achieve that $t_{msg} < 100ms$. As we can see in Table \ref{table:insert_edm}, 2500 messages are decoded and inserted into the database in $\sim 54 ms$ ($\sim 21 \mu$s per message), making the batch sizes of 2500, 5000 and 10000 not suitable for real-time use cases, but still being a good alternative for road traffic analysis use cases or other less latency dependant applications. The ETSI standard for CAM messages supports a message period of up to 1 s. So batch sizes larger than 2500 messages can be acceptable for some use cases. However, as explained previously, this work is targeting applications that require a data update frequency of 10 Hz.

Considering a time window of $50 ms$ for our $t_{buffer}$ and as our solution can insert 1000 messages into the database in $\sim 26 ms$ ($\sim 26 \mu$s per message), as long as $t_{send} < 24 ms$, which is a reasonable latency for an OBU-MEC communication in a 5G network or over IEEE 802.11p communications \cite{Fernandez2023,9798350}, we can consider that the messages are stored and available into the database in less than $100 ms$. By selecting a time window of $50ms$ and considering a normal distribution in the reception of the messages, the EDM is able to handle 2000 vehicles sending messages at 10
Hz. This time window can be easily configured to a different time span, offering less processing latency depending on the use case or expected quantity of vehicles connected to the MEC server.

\subsection{Data querying}
As it would be infeasible to assess the performance of querying the EDM in a real-world setting (requiring the deployment of thousands of vehicles), we instead conducted batch insertions to the database using a virtual testing set-up and queried the database after each insertion to measure performance. The following queries were analysed:

%1-D, 2-E, 3-A, 4-G, 5-F
\begin{itemize}
    \item Query 1: Group vehicles by ID and select the last message received from each vehicle in a given latitude and longitude region.
    
    \item Query 2: Group vehicles by ID and select the last message received from each vehicle in a given H3 index.
    
    \item Query 3: Select all messages received in the last 100ms. 
    
    \item Query 4: Select all messages received in the last 100ms in a given latitude and longitude region.
    
    \item Query 5: Select all messages received in the last 100ms in a given H3 index.
    
\end{itemize}

To facilitate an equal comparison between queries involving geoposition filters (Query 1, 2, 4 and 5), the original latitude and longitude values of the messages obtained from the SUMO simulations were modified, separating the vehicles into 20 different h3 index areas. Through these modifications, we can ensure that the number of vehicles retrieved from the queries will match the expected quantity in both geoposition filters. Regarding queries involving filters by timestamp (Query 3, 4, 5), the time window has been increased in the cases where the insertion time is longer than 100 ms to match the insertion time for that batch size. The average time obtained from performing 1000 queries for each batch size can be seen in Table \ref{table:querying}. 
\begin{table*}[]
\centering 
\caption{Mean (standard deviation) querying time in milliseconds for different batch sizes.}
\label{table:querying}
\begin{tabular}{c|c|c|c|c|c|}
\cline{2-6}
 & \textbf{Query 1} & \textbf{Query 2} & \textbf{Query 3} & \textbf{Query 4} & \textbf{Query 5} \\ \hline
\multicolumn{1}{|l|}{\textbf{Batch 100}} & 17.86 (6.66) & 17.7 (6.04) & 9.06 (1.92) & \textbf{7.26 (2.45)} & 7.36 (2.21) \\ \hline
\multicolumn{1}{|l|}{\textbf{Batch 1000}} & 62.51 (16.0) & 55.16 (13.53) & 17.61 (8.0) & 15.13 (6.61) & \textbf{10.01 (5.16)} \\ \hline
\multicolumn{1}{|l|}{\textbf{Batch 2500}} & 172.51 (33.16) & 138.3 (27.84) & 31.52 (8.77) & 24.69 (9.51) & \textbf{15.92 (8.06)} \\ \hline
\multicolumn{1}{|l|}{\textbf{Batch 5000}} & 686.36 (127.06) & 548.96 (115.52) & 82.44 (17.02) & 86.28 (12.86) & \textbf{54.15 (12.38)} \\ \hline
\multicolumn{1}{|l|}{\textbf{Batch 10000}} & 838.65 (183.57) & 728.56 (164.15) & 160.77 (39.18) & 116.36 (19.21) & \textbf{74.05 (16.44)} \\ \hline
\end{tabular}
\end{table*}

The usage of aggregation functions to group elements within the database based on ID (Query 1 and 2) exhibits inferior performance compared to the other queries presented, even when utilising geoposition filters. However, taking advantage of the temporal filtering offered by the TSDB in combination with geoposition filters (Query 4 and 5) significantly improves querying time, resulting in the most efficient performance among the presented queries, with a reduction of querying time by over 10 times in certain scenarios. Even among the geoposition queries, those utilising H3 indexing (Query 2 and 5) demonstrate superior performance compared to those that compare the latitude and longitude of the inserted values against predefined geographic regions (Query 1 and 4). 

\begin{figure}[h]
    \centering
    \includegraphics[width=0.48\textwidth]{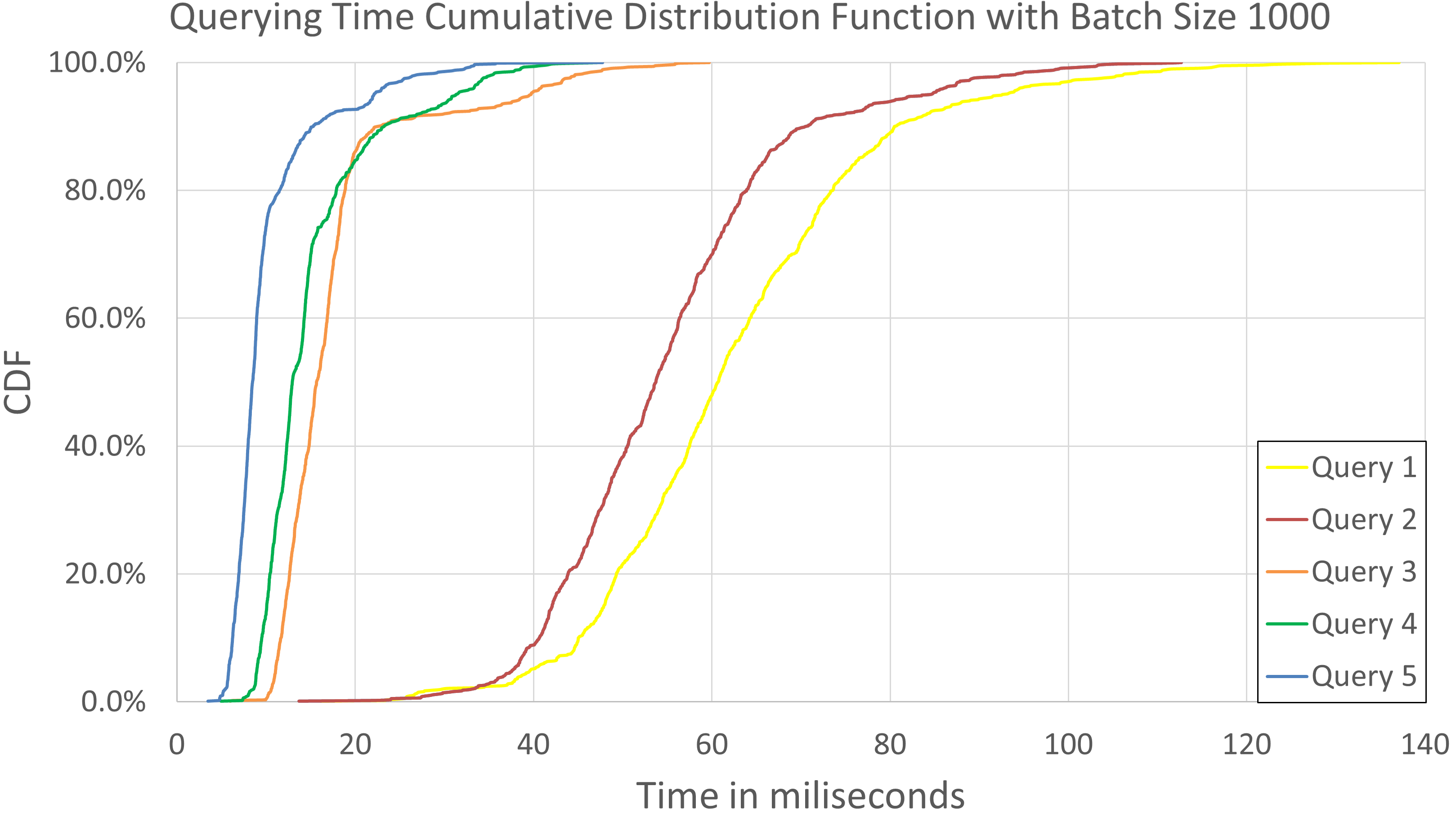}
    \caption{ Querying time CDF for the proposed queries with batch size 1000.}
    \label{fig:cdf_query}
\end{figure}

For the proposed batch size of 1000 messages, the processing time for each query can be seen in Fig. \ref{fig:cdf_query}. We can see how queries not using aggregation functions (Query 3, 4 and 5) show better performance. Specifically, for the chosen batch size, these queries achieve query results of under 20ms in $90\%$ of the cases. However $10\%$ of the queries execution time shows inferior performance, even reaching 40ms in the worst cases. We attribute these performance outliers to the low cache size of 1MB in the TSDB, which results in frequent cache clearing and increased processing times.

%In \cite{Fernandez2017}, they collect the E2E latency requirements for different road safety services, including: Road Hazard Signalling service (RHS), Longitudinal Collision Risk Warning application (LCRW) and Intersection Collision Risk Warning application (ICRW). The E2E latency requirements for these services should be less than 300ms.

\section{Conclusions}\label{sec:conclusions}
The advent of advanced cellular networks (5G and beyond) brings the possibility of generating a dynamic map in the MEC that collects and stores the dynamic information of connected vehicles. Our proposed EDM can be used in low-latency use cases by applications deployed in the MEC to, for instance, detect risk situations, assist drivers, or identify relevant traffic participants for a specific road user. Using a modest computer and the proposed architecture, each MEC can support around 2000 vehicles sending messages at 10 Hz to the MEC server. Using geoindexing approaches to filter data improves query performance significantly, enabling real-time querying to the EDM even in automotive scenarios where data querying involves 2500 vehicles. Demonstrating that the bottleneck is in the insertion rather than in querying the database. Even larger scales can be reached either by decreasing OBU-MEC communication latency (24 ms were considered in the study which is a conservative value), increasing MEC computation capability, reducing the C-ITS messaging frequency (the standard supports a range of 1-10 Hz) or testing different cache sizes for the database.

%%%%%%%%%%%%%%%%%%%%%%%%%%%%%%%%%%%%%%%%%%%%%%%%%%%%%%%%%%%%%%%%%%%%%%%%%%%%%%%%

% default value was causing issues in bibliography (-12)
\addtolength{\textheight}{0cm}   % This command serves to balance the column lengths
                                  % on the last page of the document manually. It shortens
                                  % the textheight of the last page by a suitable amount.
                                  % This command does not take effect until the next page
                                  % so it should come on the page before the last. Make
                                  % sure that you do not shorten the textheight too much.

\bibliographystyle{IEEEtran}
% argument is your BibTeX string definitions and bibliography database(s)
\bibliography{biblio}

\end{document}